\documentclass[11pt]{article}

\usepackage[utf8]{inputenc}
\usepackage[T1]{fontenc}
\usepackage{microtype}
\usepackage{fullpage}
\usepackage{quantikz}
\usepackage{mathtools}
\usepackage{amsmath}
\usepackage{amsfonts}
\usepackage{amsthm}
\usepackage{algpseudocode}
\usepackage{algorithm}
\usepackage{float}
\usepackage{graphicx}
\usepackage{hyperref}
\usepackage{enumitem}
\usepackage[table]{xcolor}
\usepackage{natbib}
\usepackage[toc,page]{appendix}
\usepackage{chngcntr}
\usepackage{titletoc}
\definecolor{citecolor}{HTML}{0071BC}
\definecolor{linkcolor}{HTML}{ED1C24}
\definecolor{mydarkblue}{rgb}{0,0.08,0.45}
\hypersetup{colorlinks=true, linkcolor=mydarkblue, citecolor=mydarkblue,urlcolor=mydarkblue}
\usepackage{smile}

\makeatletter
\@ifundefined{theorem}{\newtheorem{theorem}{Theorem}[section]}{}
\@ifundefined{lemma}{\newtheorem{lemma}[theorem]{Lemma}}{}
\@ifundefined{proposition}{\newtheorem{proposition}[theorem]{Proposition}}{}
\@ifundefined{corollary}{\newtheorem{corollary}[theorem]{Corollary}}{}
\@ifundefined{claim}{}{}
\theoremstyle{definition}
\@ifundefined{definition}{\newtheorem{definition}[theorem]{Definition}}{}
\theoremstyle{remark}
\@ifundefined{remark}{\newtheorem{remark}[theorem]{Remark}}{}
\makeatother

\usepackage{fancyhdr}
\newcommand{\firstpageheader}[1]{%
\thispagestyle{fancy}%
\fancyhf{}%
\fancyhead[L]{\small \rmfamily #1}%
\fancyfoot[C]{\thepage}%
}

\usepackage{braket}
\newcommand{\projectpage}[1]{%
\begin{center}
\small
\vspace{-1.25ex}
\texttt{Project Page}: \url{#1}
\end{center}
}

\ifdefined\final
\usepackage[disable]{todonotes}
\else
\usepackage[textsize=tiny]{todonotes}
\fi
\newcommand{\ip}[2]{\langle #1,#2\rangle}

\newcommand{\wh}{\widehat}

\newcommand{\calK}{\mathcal{K}}

\providecommand{\ZZ}{\mathbb{Z}}
\providecommand{\bm}{\boldsymbol}
\providecommand{\bbb}{\mathbf{b}}
\newcommand{\poly}{\operatorname{poly}}
\newcommand{\negl}{\operatorname{negl}}

\title{Exact Coset Sampling for Quantum Lattice Algorithms}
\author{Yifan Zhang}
\date{}

\begin{document}
\maketitle

\begin{abstract}
We revisit the post-processing phase of Chen's Karst-wave quantum lattice algorithm~\citep{chen2024quantum} in the Learning with Errors (LWE) parameter regime.
Conditioned on a transcript $E$, the post-Step~7 coordinate state on $(\ZZ_M)^n$ is supported on an affine grid line
\[
\{\,j\Delta+\bm v^\ast(E)+M_2\bm k \bmod M : j\in\ZZ,\ \bm k\in\calK\,\},
\]
with $\Delta=2D^2\bbb$, $M=2M_2=2D^2Q$, and $Q$ odd.
The amplitudes include a quadratic Karst-wave chirp $\omega_Q^{-j^2}$ and an unknown run-dependent offset $\bm v^\ast(E)$.
We show that, in the access model stated below, Chen's Steps~8--9 can be replaced by a single direct post-processing routine:
measure the deterministic residue $\tau:=X_1\bmod D^2$, obtain the run-local class $v_{1,Q}:=v_1^\ast(E)\bmod Q$ as explicit side information in our access model, apply a $v_{1,Q}$-dependent diagonal quadratic phase on $X_1$ to cancel the chirp, and then apply $\mathrm{QFT}_{\ZZ_M}^{\otimes n}$ to the coordinate registers.
The routine never needs the full offset $\bm v^\ast(E)$; its only offset input is the one-dimensional gauge residue needed to remove the chirp.
Under Additional Conditions~AC1--AC4 and the spectral concentration property AC5, a measured Fourier outcome $\bm u\in\ZZ_M^n$ satisfies the resonance
\[
\ip{\bbb}{\bm u}\equiv 0\pmod Q
\]
with probability $1-\negl(n)$.
Moreover, conditioned on resonance, the reduced outcome $\bm u\bmod Q$ is exactly uniform on the dual hyperplane
\[
H=\{\bm v\in\ZZ_Q^n:\ip{\bbb}{\bm v}\equiv 0\pmod Q\}.
\]
For factorized chirp-free loop envelopes that are shifted discrete Gaussians in the loop variable, with no residual linear phase after the exact AC4 gauge correction and with finite-window and implementation errors negligible in an explicit $Q$-grid energy sense, AC5 follows from one-dimensional Poisson-summation decay once the loop-envelope width satisfies $\sigma_J\gtrsim Q\log n$.
\end{abstract}

\firstpageheader{Preprint -- Work in Progress}

\section{Introduction}

Fourier sampling quantum algorithms for lattice problems prepare a structured superposition, and a Fourier transform then reveals modular linear structure~\citep{regev2004quantum,regev2009lattices}.
Chen's windowed quantum Fourier transform with complex Gaussian windows~\citep{chen2024quantum} fits this pattern.
The $n$ coordinate registers of the post-Step~7 state $\ket{\tilde{\varphi}_7}$ live in $\ZZ_M$, with $M=2M_2$.
The one-shot routine in Algorithm~\ref{algo:dual-sample} performs the run-local phase correction described below and then applies $\mathrm{QFT}_{\ZZ_M}^{\otimes n}$ directly to these registers.
The post-Step~7 state has affine support
\begin{equation}\label{eq:phi7-tilde}
\ket{\tilde{\varphi}_7}
= \sum_{j\in\ZZ}\ \sum_{\bm k \in \calK}\alpha_E(j,\bm k)\,
\omega_{Q}^{-j^{2}}\,
i^{\|\bm k\|^2}\,\ket{\mathbf X(j) + M_2 \bm k}_M,
\end{equation}
for an outcome-dependent amplitude profile $\alpha_E(j,\bm k)$ and an index set $\calK\subseteq\{0\}\times\{0,1\}^{n-1}$ that comes from the grid-state construction.
When it is clear from context, we suppress the transcript index~$E$.
The phase factor $i^{\|\bm k\|^2}$ comes from Chen's grid-state preparation procedure.
We view the coordinate values as elements of $(\ZZ_M)^n$ throughout.
Accordingly, $\mathbf X(j):=2D^2 j\,\bbb+\bm v^\ast$ is interpreted componentwise modulo $M$, and $\bm k$ records the $M_2$-shift term $M_2\bm k$.
Reducing the coordinate registers modulo $M_2$ eliminates this shift and yields the affine congruence
\[
\mathbf X(j)\equiv 2D^2j\,\bbb+\bm v^\ast\pmod{M_2}
\]
used in Additional Condition~AC1.

The additional factor $\omega_Q^{-j^2}=\exp(-2\pi i j^2/Q)$ is the Karst-wave chirp.
Step~$9^\dagger$ removes it using only the first coordinate register and the run-local residue $v_{1,Q}=v_1^\ast(E)\bmod Q$ promised by Additional Condition~AC4.
We denote the chirp-free state obtained after this run-local phase correction by $\ket{\varphi_7}$; it satisfies
\begin{equation}\label{eq:phi-intro}
\ket{\varphi_7}
= \sum_{j\in\ZZ}\ \sum_{\bm k \in \calK}\alpha_E(j,\bm k)\, i^{\|\bm k\|^2}\,\ket{\mathbf X(j) + M_2 \bm k}_M.
\end{equation}

Additional Condition~AC4 isolates the only extra access needed to cancel the Karst-wave chirp using only the first coordinate register:
after measuring the deterministic residue $\tau:=X_1\bmod D^2$, the post-processing must obtain the run-local class $v_{1,Q}:=v_1^\ast(E)\bmod Q$.
This information is not implied by AC1 alone; in our access model, it is provided as a short run-local gauge value.
If the gauge value is available coherently inside the preparation, one can compile a canonical-gauge variant with $\lambda(E)=0$ (Proposition~\ref{prop:AC4-gauge}), in which case $v_{1,Q}=\operatorname{ctr}(\tau)\bmod Q$.
Under Additional Condition~AC1, the first coordinate satisfies the precise congruence
\[
X_1(j)\equiv -2D^2 j + v_1^\ast(E) \pmod{M_2}
\]
for every $j$ in the effective window.
Write $\Delta:=2D^2\bbb$ and recall $M = 2M_2 = 2D^2Q$.
Since $D^2,Q\mid M$, reducing any physical basis value $x\in\ZZ_M$ modulo $D^2$ or $Q$ is unambiguous.
In particular,
\[
X_1 \bmod D^2 \equiv v_1^\ast \bmod D^2,\qquad (X_1\bmod Q)-v_{1,Q} \equiv -2D^2 j \pmod Q,
\]
where under AC4 the post-processing computes $v_{1,Q}=v_1^\ast(E)\bmod Q$ from $\tau=X_1\bmod D^2$ together with the run-local gauge value.

In Chen's post-processing, one ultimately outputs non-zero Fourier samples in $\ZZ_{M_2}^n$ whose reduction modulo $Q=M/(2D^2)$ satisfies a homogeneous dual relation.
In our direct-QFT variant, we instead measure $\bm u\in\ZZ_M^n$ and store only the reduction $\bm u_Q:=\bm u\bmod Q$.
In either view, the goal is to obtain non-zero vectors satisfying
\begin{equation}\label{eq:intended}
\ip{\bbb}{\bm u} \equiv 0 \pmod{Q},
\end{equation}
where $Q := M/(2D^2)$ is an odd integer.
The short vector $\bbb=(b_1,\dots,b_n)\in\ZZ^n$ stores the hidden data and satisfies the normalization $b_1=-1$.
The offset $\bm v^\ast$ is unknown and can change from run to run.
The algorithm then collects $O(n)$ such samples and solves the resulting linear system modulo $Q$ to recover $\bbb\bmod Q$ and hence the hidden data vector.

Chen's published Steps~8--9 enforce relations of the form Equation~\eqref{eq:intended} via additional offset-handling and a domain-extension mechanism that is tailored to a particular parameterization.
Step~$9^\dagger$ provides an alternative post-processing designed for an access model in which the affine offset $\bm v^\ast(E)$ may vary across runs and no special factorization of $Q$ is assumed.
It uses only the affine structure $\mathbf X(j)=j\Delta+\bm v^\ast$ with $\Delta=2D^2\bbb$, the normalization $b_1=-1$, a run-local chirp cancellation acting on $X_1$ (Additional Condition~AC4), and a direct $\mathrm{QFT}_{\ZZ_M}^{\otimes n}$ on the coordinate registers.
Under Additional Condition~AC5, the resulting outcomes satisfy $\ip{\bbb}{\bm u}\equiv 0\pmod Q$ except with negligible failure probability.
For the Gaussian-envelope regime satisfying the explicit hypotheses below, Proposition~\ref{prop:AC5-holds} proves AC5 from one-dimensional Fourier decay of the chirp-free loop envelope, after excluding residual linear phases and controlling truncation and approximation errors on the $Q$-point Fourier grid.

When we instantiate the front end with the $q$-ary lattice $L_q^\perp(\mathbf A)$ from~\cite{chen2024quantum}, the new Step~$9^\dagger$ recovers the planted shortest vector $\bbb=[-1,2p_1 s^\top,2p_1 e^\top]^\top$ and hence the LWE secret $s$ and error $e$ in the chosen-secret regime considered there.
Combined with the classical reductions of~\cite{chen2024quantum}, this yields a quantum algorithm for standard LWE in the same parameter regime, conditional on Additional Conditions~AC1--AC4 and the Gaussian-envelope hypotheses proving AC5.

\section{Background and Access Model}
\label{sec:prelim}

\paragraph{Notations.}
Let $q\in\mathbb N$ and write $\ZZ_q=\mathbb{Z}/q\mathbb{Z}$ with representatives in $(-q/2,q/2]$.
Vectors are written in bold.
We use $\ip{\cdot}{\cdot}$ for the inner product.
For $T\in\mathbb N$ we write $\omega_T:=\exp(2\pi i/T)$.
The physical coordinate registers live in $\ZZ_M^n$, where
\[
M := 2M_2=2D^2Q, \qquad M_2 := D^2Q.
\]
We will repeatedly reduce physical basis values $x\in\ZZ_M$ modulo $M_2$, $D^2$, or $Q$.
Since $D^2,Q\mid M_2\mid M$, the canonical reduction maps
\[
\ZZ_M \to \ZZ_{M_2},\qquad
\ZZ_M \to \ZZ_{D^2},\qquad
\ZZ_M \to \ZZ_Q
\]
are well-defined ring homomorphisms.

\paragraph{Fourier-transform convention.}
We use the plus-sign convention
\[
\mathrm{QFT}_{\ZZ_M}^{\otimes n}\ket{\bm x}
=
M^{-n/2}\sum_{\bm u\in\ZZ_M^n}\omega_M^{\ip{\bm u}{\bm x}}\ket{\bm u}.
\]
Changing to the opposite convention conjugates the phases and replaces $\bm u$ by $-\bm u$; the resonance hyperplane is unchanged.

\paragraph{Modular inner products.}
For $\bm a\in\ZZ^{n}$ and $\bm u\in(\ZZ_M)^n$ we write
\[
\ip{\bm a}{\bm u}\bmod Q \ :=\ \sum_{i=1}^n (a_i\bmod Q)\cdot (u_i\bmod Q)\ \in\ \ZZ_Q.
\]
When we write $\ip{\bm a}{\bm u}\equiv 0\pmod Q$, this is the intended meaning.

\paragraph{Representative conventions.}
We use the standard representatives in $\{0,\dots,M-1\}$ for $\ZZ_M$.
All coordinate-register kets $\ket{\cdot}_M$ are labelled by elements of $\ZZ_M$ in these representatives.
We also routinely identify $t\in\ZZ_T$ with its representative in $\{0,\dots,T-1\}$ when evaluating phases $\omega_T^t$.
When we refer to reductions modulo $M_2$, $D^2$, or $Q$, we always mean the canonical ring homomorphisms from $\ZZ_M$.

We write
\[
0\!\mid\!\ZZ^{n-1}:=\{0\}\times\ZZ^{n-1},
\qquad
0\!\mid\!\{0,1\}^{n-1}:=\{0\}\times\{0,1\}^{n-1}.
\]
Since $M_2=M/2$, the basis shift $M_2\bm k\bmod M$ depends only on $\bm k\bmod 2$ because $M_2(\bm k+2\bm e_i)\equiv M_2\bm k \pmod M$.
The phase $i^{\|\bm k\|^2}$ also depends only on $\bm k\bmod 2$.
Accordingly, throughout we fix $\calK\subseteq 0\!\mid\!\{0,1\}^{n-1}$ as a set of parity representatives.
Whenever the front end naturally produces a sum over a larger grid index set, for example $\bm k\in 0\!\mid\!\ZZ^{n-1}$ as in Chen's expression, we group all terms with the same parity class $\bm k\bmod 2$ into a single coefficient.
Thus, restricting to $\calK$ incurs no loss of generality: any dependence on $\bm k+2\ZZ^{n-1}$ is absorbed into the amplitudes $\alpha_E(j,\bm k)$.

The physical coordinate block in Chen's Step~7 lives in $\ZZ_M$ with $M=2M_2$.
When this helps to avoid confusion, we write the reduction modulo $M$ with a subscript $_M$.
For a coordinate slice we write $\bm x_{[2..n]}:=(x_2,\dots,x_n)$.
Set $\Delta := 2D^2\,\bbb \in \ZZ^n$.

All displayed state sums below omit their normalizing constants.
They should be read after collecting coincident computational-basis labels.
For algebraic manipulations it is convenient to keep the formal labels $(j,\bm k)$.
This causes no ambiguity: if two formal labels produce the same physical basis vector in $(\ZZ_M)^n$, then their amplitudes are added in that basis coefficient.
When AC5 below assumes a factorization $\alpha_E(j,\bm k)=\alpha_E(j)\beta_E(\bm k)$, the assumption is imposed on these formal coefficients after the parity regrouping of $\bm k$.
This is the correct level at which to state the hypothesis.
Indeed, the Fourier amplitude of the collected physical state is exactly the sum over formal labels, since coincident formal labels have the same computational-basis ket and therefore the same $\mathrm{QFT}_{\ZZ_M}^{\otimes n}$ phase.
Thus possible collisions in the line parametrization do not create an additional consistency condition.
In particular, because the first coordinate has $k_1=0$ and $b_1=-1$, equality of physical basis labels forces the corresponding loop labels to be congruent modulo $Q$.
Thus the chirp-correction phase $j^2\bmod Q$ and the direct-QFT phases used below are well-defined on collected physical basis states.

We use $\poly(\cdot)$ for an unspecified fixed polynomial and $\negl(n)$ for a negligible function of~$n$.
In the intended LWE instantiation (Definition~\ref{def:retuned-karst-wave-config}) we have $Q=\poly(n)$ and $Q\ge5$, hence $Q^{-(n-1)}=\negl(n)$.

\paragraph{Idealized unitaries vs.\ circuit precision.}
For clarity, we treat $\mathrm{QFT}_{\ZZ_M}$ and the controlled root-of-unity phases used in the chirp correction as exact unitaries.
In a standard uniform gate set, these operations can be implemented to precision $2^{-\poly(n)}$, which perturbs all output distributions by at most $\negl(n)$ in total variation distance.
We suppress this approximation issue throughout, since it does not affect any of the modular-algebraic arguments.
See, e.g., standard treatments of approximate QFT and fault-tolerant synthesis~\citep{nielsen2010quantum}.

Finally, we write $\operatorname{ctr}:\ZZ_{D^2}\to\ZZ$ for the centered lift:
for $\tau\in\ZZ_{D^2}$, $\operatorname{ctr}(\tau)$ is the unique integer in $(-D^2/2,D^2/2]$ that is congruent to $\tau$ modulo $D^2$.
We use $\operatorname{ctr}(\tau)$ as a canonical integer representative when converting $\tau=X_1\bmod D^2$ into arithmetic in $\ZZ_Q$.

At the frontier right before Step~8, and for fixed outcomes $E$, Lemma~\ref{lem:offset-chen} gives $\mathbf X(j)\equiv j\Delta+\bm v^\ast \pmod{M_2}$ for every $j$ in the effective window.

\paragraph{Parameter identification from \cite{chen2024quantum}.}
In Chen's notation, $M=2(t^2{+}u^2)$ and $x=D\bbb$, where $\bbb$ is the short vector from the LWE-to-lattice reduction.
We follow this identification.
Chen's Step~9 defines a relabelled vector $\bbb^\ast$.
Our post-processing never uses $\bbb^\ast$.
We keep all formulas in terms of~$\bbb$ instead.

Under Chen's parameter constraints, the quantity $M/(2D^2)$ is an odd integer.
We write
\[
Q := \frac{M}{2D^2},\qquad M_2:=\frac M2 = D^2Q.
\]
In Chen's concrete instantiation, $Q$ happens to admit a factorization into pairwise coprime odd factors; the new Step~$9^\dagger$ does not use this extra structure and only relies on $Q$ being odd.

Chen uses the letter $P$ for a large loop modulus with $P \approx M^2/2$.
We keep this convention and reserve $P$ only for that modulus.

\paragraph{Index domain vs.\ first coordinate.}
The preparation uses a loop index $j$.
This index is the discrete time variable of the Karst wave and appears in~\cite{chen2024quantum} as a summation index.
It runs over the large modulus $P$.
We denote the register by $J\in\ZZ_P$.

The first coordinate satisfies
\[
X_1\equiv 2D^2 j\,b_1+v_1^\ast\pmod{M_2}.
\]
In Chen's setup, $v_1^\ast$ depends on the measurement outcomes $(y', z', h^\star)$.
The experiment cannot be reset in a controlled way to obtain the same outcomes and the same $v^\ast$.
Thus we cannot query the state preparation on chosen basis inputs $j$ and then compute $\mathbf X(j)$ either classically or into an extra register.
We must work with the single quantum state $|\varphi_7\rangle$ that a successful run outputs.

\paragraph{Parameter re-tuning.}
As in~\cite{chen2024quantum}, we run a classical outer loop over polynomially many candidates for $\|\bbb\|^{2}$ and, for each guess, set
\[
Q=(c{+}1)\|\bbb\|^2,\qquad M=2D^2Q,\qquad M_2=D^2Q.
\]
The scaling parameter $D$ can then be chosen odd, coprime to $Q$, and $\poly(n)$-bounded so as to satisfy the geometric separation requirement of the Karst-wave front end.
The additional analytic requirement used by Step~$9^\dagger$ is the one-dimensional spectral concentration property recorded in Additional Condition~AC5.
It follows in the factorized Gaussian-envelope regime once the loop-envelope width satisfies $\sigma_J\gtrsim Q\log n$, provided the no-linear-phase and perturbation hypotheses of Proposition~\ref{prop:AC5-holds} hold.
All subsequent correctness statements are conditional on the existence of a re-tuned front-end configuration satisfying Definition~\ref{def:retuned-karst-wave-config} and Additional Conditions~AC1--AC5.
Equivalently, for the Gaussian front ends covered by Proposition~\ref{prop:AC5-holds}, the subsequent correctness statements use AC1--AC4 together with the proved spectral conclusion AC5.

\begin{definition}[Re-tuned Karst-wave configuration]
\label{def:retuned-karst-wave-config}
Take LWE parameters $(\ell,m,q,\beta)$ in the regime of~\cite{chen2024quantum} and set $n:=1+\ell+m$.
Let $L_q^\perp(\mathbf A)$, the planted shortest vector $\bbb$, and the integers $(p_\eta)_\eta=(p_1,\dots,p_\kappa)$ be as in Section~3.2 of~\cite{chen2024quantum}.
We assume $\kappa\ge 2$ and $\kappa-1\le \ell$, so that the $(\kappa-1)$ chosen secret coordinates fit in the $\ell$-dimensional secret.
A re-tuned Karst-wave configuration is any choice of front-end parameters
\[
(c,D,r,s,t,u,P,M,Q)
\]
that satisfies the following properties.
\begin{enumerate}[label=(R\arabic*),leftmargin=1.9em]
\item
The integers $p_1,\dots,p_\kappa$ are odd and pairwise coprime.
They control the chosen-secret embedding and the parsing of $\bbb$ in the LWE extraction step, but they do not impose any algebraic constraint on~$Q$.
Fix a polynomially bounded positive integer $c=c(n)\in4\mathbb Z_{>0}$ and set
\[
Q := (c{+}1)\|\bbb\|^2.
\]
In the chosen-secret LWE embedding of~\cite{chen2024quantum}, $\bbb=[-1,2p_1 s^\top,2p_1 e^\top]^\top$ has one odd coordinate and all remaining coordinates even, hence $\|\bbb\|^2\equiv 1\pmod 4$.
Since $c\in4\ZZ$, this makes $Q=(c{+}1)\|\bbb\|^2$ automatically odd.
Since $b_1=-1$, we also have $\|\bbb\|^2\ge1$, and hence $Q\ge5$.
In the LWE parameter regime under consideration $\|\bbb\|^2=\poly(n)$, so this choice gives $Q=\poly(n)$.
We also require that $Q$ satisfies
\[
\|\bbb\|_\infty \le 2p_1\beta\log n < Q/2,
\]
so that $\bbb\bmod Q$ uniquely determines $\bbb$ by centered lifting.

\item
Choose an odd scaling parameter $D$ that is coprime to $Q$ and polynomially bounded in~$n$.
Define
\[
u^2 := \|D\bbb\|^2 = D^2\|\bbb\|^2,\qquad
t^2 := c\,u^2,
\]
and set the Karst-wave moduli
\[
M := 2(t^2+u^2)=2D^2Q,\qquad M_2 := M/2 = D^2Q,\qquad P := M(t^2+u^2)=\frac{M^2}{2}.
\]
Then $2D^2$ is a unit modulo $Q$.

\item
Choose $r,s>0$ so that Chen's Karst-wave constraints C.5--C.7 hold, and the resulting geometric width parameter
\[
\sigma_{\mathrm{geom}}:=\frac{rs^{2}t}{u\sqrt{r^{4}+s^{4}}}
\]
obeys $2\log n<\sigma_{\mathrm{geom}}<D/(4\log n)$.
We require the chosen front-end parameters to satisfy these inequalities with all parameters bounded by $\poly(n)$.
This is the same geometric separation regime used to extract the centers of the Gaussian balls.

\item
For the resulting complex-Gaussian front end, let $\sigma_J$ denote the Gaussian width parameter of the one-dimensional envelope $\alpha_E(j)$ along the loop index $j$ after the AC4 chirp and gauge correction, in the convention
\[
g_\sigma(x)=\exp(-\pi x^2/\sigma^2).
\]
A configuration used by Step~$9^\dagger$ must satisfy
\[
\sigma_J \ge C_{\mathrm{spec}}\,Q\log n
\]
for some fixed absolute constant $C_{\mathrm{spec}}>0$.
A lower bound of this form keeps the main Gaussian spectral leakage in Additional Condition~AC5 bounded by $\exp(-\Omega(\log^2 n))$ when the Gaussian-envelope hypotheses of Proposition~\ref{prop:AC5-holds} hold.
Those hypotheses also require that the AC4 correction leaves no nontrivial residual linear phase in $j$, and that finite-window truncation and front-end approximation errors are negligible in the relative $Q$-grid Fourier energy sense of Proposition~\ref{prop:AC5-holds}.
The geometric width parameter $\sigma_{\mathrm{geom}}$ in~\textup{(R3)} controls the within-ball width of each extracted Gaussian ball in coordinate space; it is distinct from the loop-envelope width $\sigma_J$.
\end{enumerate}
When we refer to Chen's front end with re-tuned parameters below, we mean an outcome-conditioned circuit family compiled with a fixed choice of parameters that satisfies~\textup{(R1)}--\textup{(R4)}.
\end{definition}

\subsection{Front-end properties and additional conditions}\label{subsec:additional-conditions}

We now record the structural properties and spectral conditions that the re-tuned front end must satisfy.
These conditions allow the LWE solver in Step~$9^\dagger$ to work in the exact sampling regime.

\begin{description}
\item[AC1.]
After modulus splitting and center extraction (Steps~5--7 of~\cite{chen2024quantum}) and just before the Step~8 gadget, conditioned on any fixed transcript $E$ the post-Step~7 coordinate block is supported on
\[
\bigl\{\,j\Delta+\bm v^\ast(E)+M_2\bm k \bmod M : j\in\ZZ,\ \bm k\in\calK\,\bigr\},
\]
for some index set $\calK\subseteq 0\!\mid\!\{0,1\}^{n-1}$, where $\Delta=2D^2\bbb$ and the offset $\bm v^\ast(E)\in(\ZZ_M)^n$ depends on $E$ but not on $j$.
As above, $\calK$ is a choice of parity representatives; any larger grid index domain can be reduced to this form by regrouping all terms with the same $\bm k\bmod 2$ into a single amplitude.
Reducing modulo $M_2$ eliminates the $M_2\bm k$ term and yields the affine congruence
\[
\mathbf X(j)\equiv 2D^2j\,\bbb+\bm v^\ast(E)\pmod{M_2}.
\]
This matches the offset-coherence invariant of Chen's deferred-measurement front end (Lemma~\ref{lem:offset-chen}).
In particular, since $k_1=0$ for all $\bm k\in\calK$ and $b_1=-1$, the first coordinate satisfies
\[
X_1 \equiv -2D^2 j + v_1^\ast \pmod{M_2},\qquad
X_1 \bmod D^2 \equiv v_1^\ast \bmod D^2.
\]

\item[AC2.]
We assume throughout that $Q$ is an odd integer.
The post-processing in Step~$9^\dagger$ works for an arbitrary composite modulus $Q$.
Whenever we invert a residue class, we only do so modulo a modulus that is guaranteed to be coprime to that residue.
In particular, we do not require $Q$ to factor in any prescribed way, and the integers $(p_\eta)_\eta$ from the LWE-to-lattice reduction play no role in the algebraic structure of $Q$.

\item[AC3.]
We keep the scaling regime of~\citet{chen2024quantum}.
The parameter $D$ is odd, coprime to $Q$, and we define
\[
M := 2D^2 Q,\qquad M_2 := M/2 = D^2 Q.
\]
Since $Q$ is odd and $\gcd(D,Q)=1$, we have $\gcd(2D^2,Q)=1$, so $2D^2$ is a unit modulo $Q$.
All divisions by $D^2$ or $2D^2$ that appear in the post-processing should be read as multiplications by the corresponding inverses in $\mathbb Z_Q$; we never perform integer division by $D^2$ or $2D^2$ in $\mathbb Z$.
For the first coordinate,
\[
X_1 \equiv 2D^2 j b_1 + v_1^\ast \equiv -2D^2 j + v_1^\ast \pmod{M_2},
\]
so that reduction modulo $D^2$ kills the $j$-dependent term:
\[
X_1 \bmod D^2 \equiv v_1^\ast \bmod D^2.
\]

\item[AC4.]
Run-local chirp cancellation from the first coordinate.
The post-Step~7 amplitudes contain the quadratic chirp $\omega_Q^{-j^2}$ in the loop index $j$.
Fix a transcript $E$.
By AC1 and $b_1=-1$, for every $j$ in the effective window,
\[
X_1 \equiv -2D^2 j + v_1^\ast(E)\pmod{M_2},\qquad M_2=D^2Q.
\]
Since $k_1=0$ for all $\bm k\in\calK$, reducing modulo $D^2$ removes the $j$-term on every branch:
\[
\tau:=X_1\bmod D^2 \equiv v_1^\ast(E)\bmod D^2.
\]
Thus $\tau$ is deterministic conditioned on $E$; measuring it does not disturb the coherent superposition over $j$ and $\bm k$.
Let $\bar v_1:=\operatorname{ctr}(\tau)\in(-D^2/2,D^2/2]$.

\smallskip
\noindent\emph{Side information.}
We assume the preparation oracle also returns, as classical side information, either the residue
\[
v_{1,Q}:=v_1^\ast(E)\bmod Q\in\ZZ_Q,
\]
or equivalently a high-order gauge value $\lambda(E)\in\ZZ_Q$ satisfying
\[
v_1^\ast(E)\equiv \bar v_1+\lambda(E)D^2\pmod{M_2},
\]
so that $v_{1,Q}=(\bar v_1+\lambda(E)D^2)\bmod Q$.
See Proposition~\ref{prop:AC4-gauge} for a canonical-gauge compilation where $\lambda(E)=0$.

\smallskip
\noindent\emph{Chirp-cancellation unitary.}
Define $\mathrm{inv}:=(-2D^2)^{-1}\in\ZZ_Q$, which exists by AC2--AC3.
For each physical basis value $x\in\ZZ_M$ of $X_1$, write $x_Q:=x\bmod Q$ and define
\[
j(x)\ :=\ \mathrm{inv}\cdot\bigl((x_Q-v_{1,Q})\bmod Q\bigr)\in\ZZ_Q.
\]
We apply the diagonal unitary $U_{\mathrm{corr}}^{(v_{1,Q})}$ on $X_1$,
\[
\ket{x}_{X_1}\longmapsto \omega_Q^{\,j(x)^2}\ket{x}_{X_1},
\]
implemented by reversible arithmetic that computes $j(x)$, applies the phase, and uncomputes.
On a formal summand with loop label $j$ we have $(X_1\bmod Q)-v_{1,Q}\equiv -2D^2j\pmod Q$, hence $j(X_1)\equiv j\pmod Q$ and $j(X_1)^2\equiv j^2\pmod Q$.
If two loop labels contribute to the same physical basis vector, then the first-coordinate congruence and $b_1=-1$ force the two labels to be congruent modulo $Q$.
Thus the value of $j(X_1)^2\bmod Q$ is well-defined on the physical basis state, even when the formal line parametrization is not injective.
Therefore $U_{\mathrm{corr}}^{(v_{1,Q})}$ cancels $\omega_Q^{-j^2}$ on every branch and yields the chirp-free state $\ket{\varphi_7}$ in Equation~\eqref{eq:phi-intro}.
All arithmetic in the definition of $j(x)$ and in the square $j(x)^2$ is in $\ZZ_Q$; in particular, the phase exponent is reduced modulo $Q$.

\item[AC5.]
Spectral concentration of the chirp-free loop envelope.
Fix a transcript $E$ after the chirp cancellation of AC4.
We assume that the chirp-free amplitudes factor across the loop index and the grid label:
\[
\alpha_E(j,\bm k)=\alpha_E(j)\,\beta_E(\bm k),
\qquad \beta_E\not\equiv0 .
\]
This factorization is part of AC5 and is understood in the formal-label sense explained in Section~\ref{sec:prelim}, after the parity regrouping of $\bm k$.
Without such a factorization, the one-dimensional leakage bound below does not by itself control the full $n$-dimensional Fourier distribution.
Define, for $\bm u\in\ZZ_M^n$,
\[
S_E(\bm u)\,:=\,\sum_{j\in\mathbb Z}\sum_{\bm k \in \calK}\alpha_E(j,\bm k)\,
i^{\|\bm k\|^2}\,\omega_{M}^{\,j\ip{\bm u}{\Delta}+M_2\ip{\bm u}{\bm k}} .
\]
Up to the $\bm u$-dependent global phase $\omega_M^{\ip{\bm u}{\bm v^\ast}}$, $S_E(\bm u)$ is the unnormalized Fourier amplitude of $\ket{\bm u}$ obtained after applying $\mathrm{QFT}_{\ZZ_M}^{\otimes n}$ to the coordinate registers of $\ket{\varphi_7}$.

We view the loop index $j\in\ZZ_P$ through its integer representative on the effective window.
We extend $\alpha_E(j)$ by zero outside an interval $\{|j|\le J_{\max}\}$.
The effective window satisfies $J_{\max}=\poly(Q\log n)\ll P$ in the Karst-wave regime below; in particular, we assume $J_{\max}<P/2$ so that the embedding of $j\in\ZZ_P$ into integers is unambiguous on the support of the state.
The negligible error from truncation and front-end approximation is absorbed into the perturbation term in Proposition~\ref{prop:AC5-holds}.
Define the one-dimensional Fourier transform of the loop envelope
\[
\wh\alpha_E(\theta):=\sum_{j\in\ZZ}\alpha_E(j)\,e^{2\pi i j\theta}.
\]
All transcript-dependent linear phase remaining after AC4 is included in $\alpha_E$.
AC5 requires that, after the AC4 gauge correction, the dominant $Q$-grid Fourier class is the zero class.
A residual factor $e^{2\pi i\nu_E j}$ with $\nu_E\not\equiv0\pmod1$ generally shifts the dominant class and invalidates resonance at $t=0$.
Equivalently, for $t\in\ZZ_Q$ and the periodization
\[
A_E(r):=\sum_{\substack{j\in\ZZ\\ j\equiv r\!\!\!\pmod Q}}\alpha_E(j),\qquad r\in\ZZ_Q ,
\]
we have
\[
\wh\alpha_E(t/Q)=\sum_{r\in\ZZ_Q}A_E(r)\omega_Q^{rt}.
\]
A good transcript is therefore one for which the $Q$-point discrete Fourier transform of the periodized loop envelope is concentrated at the zero character.
This is the exact spectral form needed by the direct-QFT proof; no pointwise lower bound on the grid factor in the $\bm k$ variable is required.
A transcript $E$ is called good if the off-resonant Fourier mass on the $Q$-point grid $\{t/Q:t\in\ZZ_Q\}$ is negligible:
\begin{equation}\label{eq:AC5-leakage}
\sum_{t\in\ZZ_Q\setminus\{0\}}\bigl|\wh\alpha_E(t/Q)\bigr|^2
\le \negl(n)\cdot\sum_{t\in\ZZ_Q}\bigl|\wh\alpha_E(t/Q)\bigr|^2.
\end{equation}
The denominator in~\eqref{eq:AC5-leakage} is required to be nonzero.
This condition is best understood through periodization modulo $Q$.
Since $\alpha_E$ is finitely supported after truncation, Parseval on $\ZZ_Q$ gives
\[
\sum_{t\in\ZZ_Q}\bigl|\wh\alpha_E(t/Q)\bigr|^2
=
Q\sum_{r\in\ZZ_Q}|A_E(r)|^2 .
\]
Thus the denominator is nonzero exactly when the $Q$-periodized loop envelope is not identically zero.
This holds, with quantitative stability, for the Gaussian envelopes in Proposition~\ref{prop:AC5-holds} and for the concrete truncation regime of Lemma~\ref{lem:AC5-perturbation}.
We assume that a transcript drawn from the front end is good except with probability $\negl(n)$ over the front-end randomness; in the Gaussian-envelope regime this follows from Proposition~\ref{prop:AC5-holds}.
\end{description}

\paragraph{Reliability of the additional conditions.}
Additional Conditions~AC2--AC3 are enforced directly by Definition~\ref{def:retuned-karst-wave-config}.
Additional Condition~AC1 is the standard offset-coherence invariant of Chen's deferred-measurement front end (Lemma~\ref{lem:offset-chen}).
Additional Condition~AC4 is the only genuinely new access-model requirement used by Step~$9^\dagger$: it asks for run-local access to $v_{1,Q}(E)$, equivalently to a gauge value $\lambda(E)$, so that the quadratic chirp can be cancelled using only $X_1$.
This information is not implied by AC1 and is treated as explicit side information returned by the oracle.
If the gauge value is available coherently inside the preparation, one can also compile a canonical gauge via an $E$-controlled basis translation on $X_1$ (Proposition~\ref{prop:AC4-gauge}).
Finally, Additional Condition~AC5 is the one-dimensional spectral estimate needed by the direct-QFT step.
Proposition~\ref{prop:AC5-holds} proves AC5 for chirp-free factorized Gaussian loop envelopes with no residual linear phase and with negligible relative perturbation on the $Q$-point Fourier grid.
Lemma~\ref{lem:AC5-perturbation} gives a simple $\ell^1$ sufficient condition for the perturbation bound, covering standard Gaussian tail truncation at radius $\Omega(\sigma_J\log n)$.
Thus, in the Gaussian regime, AC5 is not an independent algebraic promise; it is the spectral conclusion supplied by that proposition.
Outside this Gaussian regime, AC5 remains the explicit spectral condition required by the direct-QFT analysis.

\begin{lemma}[Uniformity modulo $Q$ on the dual hyperplane]\label{lem:uniform-modQ}
Assume AC1--AC4 and fix any transcript $E$.
Let $\mu_E$ be the distribution of $\bm u\in\ZZ_M^n$ obtained by applying $\mathrm{QFT}_{\ZZ_M}^{\otimes n}$ to the coordinate registers of $\ket{\varphi_7}$ and measuring.
Define the resonance event and dual hyperplane
\[
R := \{\bm u\in\ZZ_M^n : \ip{\bbb}{\bm u}\equiv 0 \pmod Q\},\qquad H := \{\bm v\in\ZZ_Q^n : \ip{\bbb}{\bm v}\equiv 0 \pmod Q\}.
\]
If $\Pr_{\mu_E}[R]>0$, then conditioned on $R$ the reduced outcome $\bm u\bmod Q\in\ZZ_Q^n$ is exactly uniform on $H$.
In particular,
\[
\Pr[\bm u\bmod Q=\bm 0\mid R]=1/|H|=Q^{-(n-1)}.
\]
\end{lemma}

\begin{proof}
Fix $E$.
Apply $\mathrm{QFT}_{\ZZ_M}^{\otimes n}$ to the chirp-free state~\eqref{eq:phi-intro}.
For each $\bm u\in\ZZ_M^n$, the unnormalized amplitude equals
\[
\omega_M^{\ip{\bm u}{\bm v^\ast}}\cdot
\sum_{j\in\ZZ}\sum_{\bm k\in\calK}\alpha_E(j,\bm k)\,i^{\|\bm k\|^2}\,
\omega_M^{\,j\ip{\bm u}{\Delta}+M_2\ip{\bm u}{\bm k}},
\]
so the outcome probabilities are proportional to the squared magnitude of the inner sum.

Because $M_2=M/2$, we have $\omega_M^{M_2\ip{\bm u}{\bm k}}=(-1)^{\ip{\bm u}{\bm k}}$, which depends only on the parity pattern of $\bm u_{[2..n]}$ since $k_1=0$ for all $\bm k\in\calK$.
Let $\bm u_Q:=\bm u\bmod Q\in\ZZ_Q^n$ and set $t(\bm u):=\ip{\bbb}{\bm u_Q}\in\ZZ_Q$.
We can reduce $\bm u$ modulo $Q$ inside the phase because of the special form of $\Delta$.
Write $\bm u=\bm u_Q+Q\bm t'$ with $\bm t'\in\ZZ^n$ coordinatewise.
Since $\Delta=2D^2\bbb$ and $M=2D^2Q$, we have
\[
\ip{Q\bm t'}{\Delta}=Q\cdot 2D^2\ip{\bm t'}{\bbb}=M\cdot \ip{\bm t'}{\bbb}\in M\ZZ,
\]
so $\omega_M^{\,j\ip{Q\bm t'}{\Delta}}=1$ for all integers $j$.
Therefore
\[
\omega_M^{\,j\ip{\bm u}{\Delta}}
=\omega_M^{\,j\ip{\bm u_Q}{\Delta}}
=\exp\!\left(2\pi i\,j\,\frac{t(\bm u)}{Q}\right).
\]
On the event $R$ we have $t(\bm u)=0$, hence $\omega_M^{j\ip{\bm u}{\Delta}}=1$ for all $j$.
Therefore, conditioned on $R$, the outcome probabilities depend on $\bm u$ only through the parity pattern of $\bm u_{[2..n]}$.
Write this parity-dependent squared amplitude as $\Gamma_E(\rho)$, where $\rho\in\{0,1\}^{n-1}$.

Now fix any $\bm v\in H$.
The fibre $F_{\bm v}:=\{\bm u\in\ZZ_M^n:\bm u\bmod Q=\bm v\}$ consists of the vectors $\bm u=\bm v+Q\bm t'$ with $\bm t'\in\ZZ_{M/Q}^n=\ZZ_{2D^2}^n$.
Since $Q$ is odd, as $\bm t'$ ranges over $\ZZ_{2D^2}^n$ the parity pattern of $(\bm v+Q\bm t')_{[2..n]}$ is exactly uniform over $\{0,1\}^{n-1}$, independently of $\bm v$.
More precisely, every parity pattern occurs exactly $(D^2)^{n-1}$ times in the last $n-1$ coordinates, and the first coordinate contributes the common multiplicity $2D^2$.
Thus
\[
\sum_{\bm u\in F_{\bm v}}\Gamma_E(\bm u_{[2..n]}\bmod 2)
=(2D^2)(D^2)^{n-1}\sum_{\rho\in\{0,1\}^{n-1}}\Gamma_E(\rho),
\]
which is independent of $\bm v$.
Hence $\mu_E[F_{\bm v}\mid R]$ is constant over $\bm v\in H$.
This proves uniformity of $\bm u\bmod Q$ on $H$ conditioned on $R$.
Finally, because $b_1=-1$ is a unit in $\ZZ_Q$, $|H|=Q^{n-1}$, giving the stated probability of the zero vector.
\end{proof}

\begin{proposition}[Gauge side information and canonical-gauge compilation]\label{prop:AC4-gauge}
Assume AC1 and suppose the state-preparation procedure is given as an explicit circuit or oracle that outputs the coordinate registers together with a classical side-information register $E$ from which the gauge parameter $\lambda(E)\in\ZZ_Q$ of AC4, equivalently $v_{1,Q}=v_1^\ast(E)\bmod Q$, can be obtained.
Then Additional Condition~AC4 holds: the post-processing can compute $v_{1,Q}$ for the current run and implement the chirp-cancellation unitary on $X_1$.
Moreover, if $E$ is available coherently during the preparation, one can compile the preparation by appending an $E$-controlled basis translation on $X_1$ by $-\lambda(E)D^2$ modulo $M$, so that the resulting state satisfies the canonical gauge $\lambda(E)=0$.
If the preparation is a black-box state oracle that outputs only the coordinate registers and does not expose $\lambda(E)$ or $v_1^\ast(E)\bmod Q$ as side information, then AC4 becomes an additional oracle promise and is not implied by AC1 alone.
\end{proposition}

\begin{proof}
Fix a transcript $E$.
By AC1 we have $X_1\equiv -2D^2j+v_1^\ast(E)\pmod{M_2}$ with $M_2=D^2Q$.
Let $\tau(E):=v_1^\ast(E)\bmod D^2$ and let $\bar v_1:=\operatorname{ctr}(\tau(E))\in(-D^2/2,D^2/2]$.
Because $D^2\mid M_2$, there exists a unique $\lambda(E)\in\ZZ_Q$ such that
\[
v_1^\ast(E)\equiv \bar v_1+\lambda(E)D^2 \pmod{M_2}.
\]
By assumption, the post-processing can obtain $\lambda(E)$, or directly $v_{1,Q}$, from the side-information register $E$.
Hence it can recover the needed residue
\[
v_{1,Q}=v_1^\ast(E)\bmod Q=(\bar v_1+\lambda(E)D^2)\bmod Q
\]
and implement the chirp correction of AC4.
If one prefers to avoid carrying $\lambda(E)$ into the post-processing, one can instead compile the preparation itself:
inside the preparation unitary, where $E$ is available coherently, apply the controlled basis translation
\[
\ket{x}_{X_1}\mapsto \ket{x-\lambda(E)D^2 \bmod M}_{X_1}.
\]
This map is a permutation of the computational basis of $X_1$ and hence unitary.
It leaves $\tau=X_1\bmod D^2$ invariant, and it replaces the first offset modulo $M_2$ by $\bar v_1$, enforcing the canonical gauge $\lambda(E)=0$.
Here ``canonical gauge'' is a statement modulo $M_2=D^2Q$.
The representative of the first offset modulo $M$ may still differ from the centered integer $\bar v_1$ by an $M_2$ shift, but this has no effect on the residue $v_{1,Q}$ or on the subsequent Fourier analysis.
\end{proof}

\begin{lemma}[Center-referenced chirp cancellation]\label{lem:chirp-cancellation}
Assume AC1--AC4.
Then the procedure in AC4 defines, for each run-local residue $v_{1,Q}=v_1^\ast(E)\bmod Q$, a unitary $U_{\mathrm{corr}}^{(v_{1,Q})}$ on the first coordinate register such that $|\varphi_7\rangle :=
U_{\mathrm{corr}}^{(v_{1,Q})}|\tilde\varphi_7\rangle$ has the form Equation~\eqref{eq:phi-intro}.
The quadratic chirp $e^{-2\pi i j^2/Q}$ in the loop index is removed on every branch.
\end{lemma}

\begin{proof}
Under AC1, we have $X_1\equiv-2D^2 j+v_1^\ast\pmod{M_2}$ with $v_1^\ast$ independent of $j$.
Conditioned on the transcript $E$, the measurement of $\tau:=X_1\bmod D^2$ is deterministic and yields $\tau\equiv v_1^\ast\bmod D^2$ on every branch.
By AC4, the post-processing obtains $v_{1,Q}=v_1^\ast(E)\bmod Q$, hence $(X_1\bmod Q)-v_{1,Q}\equiv -2D^2j\pmod Q$ on every branch.
The reversible computation of $j(X_1)$ in AC4 is well-defined modulo $Q$ by AC2--AC3.
For $x=X_1$ on a branch with loop label $j$ we have $j(X_1)\equiv j\pmod Q$.
If several formal loop labels represent the same physical coordinate, the congruence in the first coordinate forces them to be congruent modulo $Q$, so the correction phase is the same for all such contributions.
The unitary $U_{\mathrm{corr}}$ multiplies that branch by $\exp(2\pi i j^2/Q)$.
It cancels the original chirp $e^{-2\pi i j^2/Q}$.
The map is diagonal in the computational basis of $X_1$, hence unitary.
\end{proof}

\begin{remark}[Justification of Additional Condition AC5]\label{rem:AC5-justification}
Additional Condition~AC5 is exactly the spectral statement used by Step~$9^\dagger$:
after chirp cancellation, the one-dimensional Karst-wave envelope has negligible Fourier mass on the nonzero points of the $Q$-point grid.
For the Gaussian loop envelopes in the re-tuned regime, this follows from Poisson summation and the width condition $\sigma_J\gtrsim Q\log n$, provided the AC4 correction removes every residual linear phase and the finite-window and implementation errors are negligible in the relative grid-energy sense of Proposition~\ref{prop:AC5-holds}.

In the factorized Karst-wave regime used in Proposition~\ref{prop:AC5-holds}, conditioned on a transcript $E$, the post-Step~7 amplitudes factor as $\alpha_E(j,\bm k)=\alpha_E(j)\beta_E(\bm k)$.
After chirp cancellation, the Fourier amplitude at $\bm u\in\ZZ_M^n$ factorizes as
\[
S_E(\bm u)=\wh\alpha_E(t(\bm u)/Q)\cdot G_E(\bm u),
\]
where $G_E(\bm u)$ collects the grid contribution.
Here
\[
t(\bm u)\ :=\ \ip{\bbb}{\bm u}\bmod Q\ \in\ \ZZ_Q,
\]
and $t(\bm u)$ is identified with its standard representative in $\{0,\dots,Q-1\}$ when used as an argument of $\wh\alpha_E(\cdot)$ on $[0,1)$.
Taking the envelope width $\sigma_J\gtrsim Q\log n$ yields
\[
|\wh\alpha_E(t/Q)|\le \exp(-\Omega(\log^2 n))\cdot|\wh\alpha_E(0)|
\]
for the main Gaussian term and all integers $t\not\equiv 0\pmod Q$ under the Gaussian-envelope hypotheses of Proposition~\ref{prop:AC5-holds}, by standard Poisson-summation and discrete-Gaussian Fourier-decay estimates~\citep{micciancio2007worst,peikert2010efficient,regev2010learning}.
The relative perturbation hypothesis in Proposition~\ref{prop:AC5-holds}, and concretely the $\ell^1$ criterion in Lemma~\ref{lem:AC5-perturbation}, then transfers this estimate from the ideal Gaussian to the actual front-end envelope and gives the leakage bound~\eqref{eq:AC5-leakage}.

An integer time shift $j\mapsto j-j_0$ multiplies $\wh\alpha_E(\theta)$ by a unit-modulus factor and does not affect its magnitude.
More generally, Proposition~\ref{prop:AC5-holds} allows a real Gaussian center $j_0\in\mathbb R$; the Poisson formula contains additional unit-modulus phases, but the same off-grid decay bound holds.
In contrast, a nontrivial linear phase $\alpha_E(j)\mapsto \alpha_E(j)e^{2\pi i \nu j}$ shifts the spectrum $\wh\alpha_E(\theta)\mapsto \wh\alpha_E(\theta+\nu)$ and can move the dominant residue class away from $t=0$ on the grid $\{t/Q\}$.
In our setting, if the post-processing were to ignore the gauge term $\lambda(E)D^2$ and use $v_{1,Q}=\bar v_1\bmod Q$ in place of the correct value $v_{1,Q}=(\bar v_1+\lambda(E)D^2)\bmod Q$, then the chirp-cancellation unitary would leave a residual linear phase $\omega_Q^{-\lambda(E)j}$, up to a global constant, on the loop index $j$.
This would shift the dominant residue class away from $t=0$ on the $Q$-point grid.
Additional Condition~AC4, or canonical-gauge compilation, rules out this mismatch by requiring the exact run-local residue $v_{1,Q}$.
Indeed, using $\bar v_1$ instead of the correct residue gives
\[
j_{\mathrm{wrong}}(X_1)
\equiv (-2D^2)^{-1}\bigl(-2D^2j+\lambda(E)D^2\bigr)
\equiv j-2^{-1}\lambda(E) \pmod Q,
\]
where $2^{-1}$ is taken in $\ZZ_Q$.
The correction phase is then $\omega_Q^{(j-2^{-1}\lambda(E))^2}$, and after multiplying by the original $\omega_Q^{-j^2}$ the remaining $j$-dependent factor is $\omega_Q^{-\lambda(E)j}$, up to a global constant.

No additional near-uniformity assumption is needed for the reduced samples:
Lemma~\ref{lem:uniform-modQ} shows that conditioned on $\ip{\bbb}{\bm u}\equiv 0\pmod Q$, the reduction $\bm u\bmod Q$ is exactly uniform on the dual hyperplane $H$.
\end{remark}

\begin{proposition}[Gaussian-envelope proof of AC5]\label{prop:AC5-holds}
Fix a transcript $E$.
Assume the chirp has been cancelled as in AC4 and that the amplitudes factor as
\[
\alpha_E(j,\bm k)=\alpha_E(j)\beta_E(\bm k),
\qquad \beta_E\not\equiv0 .
\]
Suppose moreover that, after zero extension to an $\ell^1(\ZZ)$ sequence, the loop envelope admits a decomposition
\[
\alpha_E(j)=c_E\,g_{\sigma_J}(j-j_0)+\varepsilon_E(j),
\qquad
g_{\sigma}(x):=\exp(-\pi x^2/\sigma^2),
\]
for some $c_E\neq0$ and $j_0\in\mathbb R$.
Here $\sigma_J$ is the Gaussian width parameter, not the standard deviation.
The main term has no residual nontrivial linear phase: all transcript dependence in the main term is contained in $c_E$ and $j_0$, and there is no additional factor $e^{2\pi i\nu_E j}$ with $\nu_E\not\equiv0\pmod1$.
Such a factor would replace $\widehat g_{\sigma_J,j_0}(\theta)$ by $\widehat g_{\sigma_J,j_0}(\theta+\nu_E)$ and would generally move the dominant $Q$-grid class away from $0$.

Let
\[
\widehat g_{\sigma,j_0}(\theta):=\sum_{j\in\ZZ}g_{\sigma}(j-j_0)e^{2\pi i j\theta}.
\]
Assume that the perturbation is negligible in relative energy on the $Q$-point Fourier grid:
\begin{equation}\label{eq:relative-grid-perturbation}
\sum_{t\in\ZZ_Q}|\widehat{\varepsilon}_E(t/Q)|^2
\le \negl(n)\cdot
\sum_{t\in\ZZ_Q}|c_E\,\widehat{g}_{\sigma_J,j_0}(t/Q)|^2 .
\end{equation}
If $Q=\poly(n)$ and $\sigma_J\ge C_{\mathrm{spec}}Q\log n$ for a fixed positive absolute constant $C_{\mathrm{spec}}$, then $E$ is good in the sense of AC5.
Consequently, any front-end family for which these hypotheses fail only with probability $\negl(n)$ satisfies AC5.
\end{proposition}

\begin{proof}
Write
\[
G_t:=c_E\,\widehat g_{\sigma_J,j_0}(t/Q),\qquad
E_t:=\widehat{\varepsilon}_E(t/Q),\qquad t\in\ZZ_Q .
\]
It suffices to prove that the nonzero-grid energy of $G=(G_t)_{t\in\ZZ_Q}$ is negligible compared with its total grid energy, and then show that the perturbation $E=(E_t)_{t\in\ZZ_Q}$ cannot destroy this ratio.

Poisson summation gives
\[
\widehat g_{\sigma,j_0}(\theta)
=
\sigma\sum_{m\in\ZZ}
\exp\!\bigl(-\pi\sigma^2(m-\theta)^2\bigr)
\exp\!\bigl(-2\pi i j_0(m-\theta)\bigr).
\]
The phase depending on $j_0$ has unit modulus.
For $\sigma\ge2$, the zero-frequency value is bounded below by the dominant $m=0$ term:
\[
|\widehat g_{\sigma,j_0}(0)|
\ge
\sigma\left(1-2\sum_{m\ge1}e^{-\pi\sigma^2m^2}\right)
\ge \sigma/2 .
\]
For $\theta\in\mathbb R$, let $\|\theta\|_{\mathbb R/\ZZ}$ denote its distance to the nearest integer.
The same Poisson formula gives absolute constants $C_0,c_0>0$ such that, for all $\sigma\ge2$,
\[
|\widehat g_{\sigma,j_0}(\theta)|
\le
C_0\sigma\exp\!\bigl(-c_0\sigma^2\|\theta\|_{\mathbb R/\ZZ}^2\bigr)
\qquad(\theta\in\mathbb R).
\]
Indeed, after translating $\theta$ by an integer, write $\delta=\|\theta\|_{\mathbb R/\ZZ}\in[0,1/2]$.
The nearest Poisson term contributes at most $\sigma e^{-\pi\sigma^2\delta^2}$, while all other terms are bounded by a Gaussian tail and are absorbed into the same expression after decreasing the absolute constant $c_0$.
If $t\not\equiv0\pmod Q$, then $\|t/Q\|_{\mathbb R/\ZZ}\ge1/Q$.
Hence
\[
\rho_G:=
\frac{\sum_{t\neq0}|G_t|^2}
     {\sum_{t\in\ZZ_Q}|G_t|^2}
\le
4C_0^2 Q\exp\!\bigl(-2c_0\sigma_J^2/Q^2\bigr)
=\negl(n),
\]
because $Q=\poly(n)$ and $\sigma_J\ge C_{\mathrm{spec}}Q\log n$.
In particular, the Gaussian grid-energy denominator is nonzero.
Moreover $\sqrt{\rho_G}=\negl(n)$, since the displayed bound is $\exp(-\Omega(\log^2 n))$ up to a polynomial factor.

Now incorporate the perturbation.
Let
\[
\eta_G:=\sqrt{\rho_G},
\qquad
\delta_{\mathrm{pert}}:=
\frac{\sum_t|E_t|^2}{\sum_t|G_t|^2}.
\]
The first part gives $\eta_G=\negl(n)$, and the perturbation hypothesis~\eqref{eq:relative-grid-perturbation} gives $\delta_{\mathrm{pert}}=\negl(n)$.
For all sufficiently large $n$, $\sqrt{\delta_{\mathrm{pert}}}<1/2$, so by the triangle inequality,
\[
\frac{\left(\sum_{t\neq0}|G_t+E_t|^2\right)^{1/2}}
     {\left(\sum_t|G_t+E_t|^2\right)^{1/2}}
\le
\frac{\eta_G+\sqrt{\delta_{\mathrm{pert}}}}{1-\sqrt{\delta_{\mathrm{pert}}}}
=\negl(n).
\]
The denominator is nonzero because
\[
\left(\sum_t|G_t+E_t|^2\right)^{1/2}
\ge
(1-\sqrt{\delta_{\mathrm{pert}}})\left(\sum_t|G_t|^2\right)^{1/2}>0 .
\]
Squaring this inequality gives exactly the leakage bound~\eqref{eq:AC5-leakage} for $\alpha_E$.
\end{proof}

\begin{lemma}[Concrete perturbation criterion on the $Q$-grid]\label{lem:AC5-perturbation}
In the setting of Proposition~\ref{prop:AC5-holds}, assume $\sigma_J\ge2$.
If
\[
\|\varepsilon_E\|_{\ell^1(\ZZ)}
\le
\xi_E\,\frac{|c_E|\sigma_J}{\sqrt Q},
\]
then
\[
\sum_{t\in\ZZ_Q}|\widehat{\varepsilon}_E(t/Q)|^2
\le
4\xi_E^2\,
\sum_{t\in\ZZ_Q}|c_E\,\widehat g_{\sigma_J,j_0}(t/Q)|^2 .
\]
Consequently, if $\xi_E=\negl(n)$, then the relative grid-energy perturbation hypothesis~\eqref{eq:relative-grid-perturbation} holds.
In particular, truncating the Gaussian main term outside an interval
\[
|j-j_0|\le L,\qquad L\ge C_{\mathrm{tail}}\sigma_J\log n,
\]
contributes a negligible relative grid perturbation for $Q=\poly(n)$, and the same remains true after adding any implementation error whose $\ell^1$ norm is at most $\negl(n)\cdot |c_E|\sigma_J/\sqrt Q$.
\end{lemma}

\begin{proof}
For every $\theta$,
\[
|\widehat{\varepsilon}_E(\theta)|
\le
\|\varepsilon_E\|_{\ell^1(\ZZ)} .
\]
Therefore
\[
\sum_{t\in\ZZ_Q}|\widehat{\varepsilon}_E(t/Q)|^2
\le
Q\|\varepsilon_E\|_{\ell^1(\ZZ)}^2
\le
\xi_E^2 |c_E|^2\sigma_J^2 .
\]
On the other hand, the proof of Proposition~\ref{prop:AC5-holds} gives
\[
\sum_{t\in\ZZ_Q}|c_E\,\widehat g_{\sigma_J,j_0}(t/Q)|^2
\ge
|c_E|^2|\widehat g_{\sigma_J,j_0}(0)|^2
\ge
|c_E|^2\sigma_J^2/4 .
\]
This proves the first claim.

For the truncation claim, the standard Gaussian tail bound gives, for $L\ge\sigma_J$,
\[
\sum_{|j-j_0|>L}g_{\sigma_J}(j-j_0)
\le
C\,\sigma_J\exp\!\left(-c\,L^2/\sigma_J^2\right)
\]
for absolute constants $C,c>0$.
If $L\ge C_{\mathrm{tail}}\sigma_J\log n$ and $Q=\poly(n)$, then
\[
\sqrt Q\exp\!\left(-c\,L^2/\sigma_J^2\right)=\negl(n),
\]
after increasing $C_{\mathrm{tail}}$ if necessary.
Thus the truncation error satisfies the displayed $\ell^1$ criterion with $\xi_E=\negl(n)$.
The implementation-error statement follows by the triangle inequality.
\end{proof}

\subsection{Access model}
\label{subsec:access-model}

\begin{definition}[Access model]\label{def:hidden-slope}
We assume circuit-level quantum access to an outcome-conditioned state-preparation procedure that, on each invocation, outputs one copy of the post-Step~7 state $|\tilde\varphi_7\rangle$ on coordinate registers
$X\in(\ZZ_M)^n$, together with a classical side-information register $E$.
The register $E$ may include Chen's measured strings and, for AC4, a short run-local gauge value sufficient to obtain the residue $v_{1,Q}:=v_1^\ast(E)\bmod Q$, equivalently $\lambda(E)\in\ZZ_Q$.
We treat $E$ as a classical output register, i.e., returned in the computational basis.
Equivalently, the oracle may measure $E$ before returning it.
This ensures that reading $E$ and any deterministic function of $E$ is compatible with maintaining coherence over the loop index $j$ and grid label $\bm k$ in $|\tilde\varphi_7\rangle$.
We do not assume this gauge value is derivable from the Step~7 coordinate registers alone; it is treated as explicit side information returned by the oracle.
Since it is fixed conditioned on $E$, it can be measured without disturbing the superposition over $j$ and $\bm k$.
As a special case, if the known preparation circuit is compiled into the canonical gauge $\lambda(E)=0$ (Proposition~\ref{prop:AC4-gauge}), then $v_1^\ast(E)\bmod Q$ can be recovered from $\tau=X_1\bmod D^2$ alone and no explicit transcript access is needed for chirp cancellation.

For each invocation there exists a transcript $E$ such that, conditioned on $E$, the reduced state on $X$ satisfies Additional Condition~AC1 with offset $\bm v^\ast(E)$ and amplitudes $\alpha_E(j,\bm k)$.
Across invocations, the transcript is re-sampled, so the offset $\bm v^\ast(E)$ may change from run to run and cannot be forced to repeat.

The post-processing is allowed to apply any quantum circuit of size $\operatorname{poly}(n)$ to the output registers, including arithmetic in $\ZZ_M$ and $\ZZ_Q$, intermediate measurements, and $\mathrm{QFT}_{\ZZ_M}^{\otimes n}$.
We do not assume the ability to query the preparation on chosen basis inputs $j$, to classically compute the full offset $\bm v^\ast(E)$, or to obtain two copies of $|\tilde\varphi_7\rangle$ with the same offset.
\end{definition}

In this model, an algorithm acts on the superposition directly.
The offset $\bm v^\ast(E)$ depends on the run-specific outcomes $E$ and behaves like a one-time pad on the coordinate values in the computational basis.
Step~$9^\dagger$ shows that this offset turns into a harmless phase pattern after a Fourier transform.
We can then sample from the dual lattice without ever learning $\Delta$ or the full offset $\bm v^\ast$ as explicit classical data.
The only offset information used by the post-processing is the single run-local residue $v_{1,Q}=v_1^\ast(E)\bmod Q$ required for chirp cancellation.

\begin{lemma}[Offset coherence]\label{lem:offset-chen}
In the deferred-measurement unitary of~\citet{chen2024quantum}, take the program point just before Step~8 and after modulus splitting and center extraction (Steps~5--7).
For each fixed transcript $E$, the post-Step~7 coordinate block in $(\ZZ_M)^n$ is supported on
\[
\bigl\{\,j\Delta+\bm v^\ast(E)+M_2\bm k \bmod M : j\in\ZZ,\ \bm k\in\calK\,\bigr\},
\]
for some index set $\calK\subseteq 0\!\mid\!\{0,1\}^{n-1}$, where $\Delta=2D^2\bbb$ and the offset $\bm v^\ast(E)\in(\ZZ_M)^n$ depends on $E$ but not on $j$.
Reducing modulo $M_2$ eliminates the $M_2\bm k$ term and yields
\[
\mathbf X(j)\equiv 2D^2j\,\bbb+\bm v^\ast(E)\pmod{M_2}.
\]
\end{lemma}

\begin{proof}
This is immediate from Chen's post-Step~7 computational-basis form:
the basis values are $2Dj\,\bm x+\bm v'+\tfrac{M}{2}\bm k\bmod M$ with $\bm k\in 0\!\mid\!\ZZ^{n-1}$ and $\bm x=D\bbb$.
Since $M_2=M/2$, only $\bm k\bmod 2$ affects $\tfrac{M}{2}\bm k\bmod M$ and the phase $i^{\|\bm k\|^2}$, so one may fix $\bm k\in 0\!\mid\!\{0,1\}^{n-1}$ as representatives.
Substituting $\bm x=D\bbb$ gives $2Dj\,\bm x=2D^2j\,\bbb=j\Delta$ with $\Delta=2D^2\bbb$, and writing $\bm v^\ast(E):=\bm v'\bmod M$ yields the claimed support in $(\ZZ_M)^n$.
Reducing modulo $M_2=M/2$ eliminates the $M_2\bm k$ term, giving $\mathbf X(j)\equiv 2D^2j\,\bbb+\bm v^\ast(E)\pmod{M_2}$.
\end{proof}

\begin{algorithm}[H]
\caption{\textsc{Step9DaggerSample}: chirp correction and direct Fourier sampling}
\label{algo:dual-sample}
\begin{algorithmic}[1]
\State \textbf{Input:} Parameters $(D,M,Q)$ and $\mathrm{inv}:=(-2D^2)^{-1}\in\ZZ_Q$.
\State \textbf{Oracle:} One copy of $|\tilde{\varphi}_7\rangle$ on coordinate registers $X\in(\ZZ_M)^n$ and side information sufficient to obtain $v_{1,Q}:=v_1^\ast(E)\bmod Q$.
\State \textbf{Output:} A reduced sample $\bm u_Q\in\ZZ_Q^n$.
\State Query the state-preparation oracle to obtain $|\tilde{\varphi}_7\rangle$ on $X$ and run-local side information $E$.
\State Measure $\tau \gets X_1 \bmod D^2$ and set $\bar v_1\gets \operatorname{ctr}(\tau)$.
\State Using $(E,\bar v_1)$, compute the run-local residue $v_{1,Q}:=v_1^\ast(E)\bmod Q$ as in AC4:
\State \hspace{1.6em}either read $v_{1,Q}$ directly from $E$, or read $\lambda(E)\in\ZZ_Q$ from $E$ and set $v_{1,Q}\gets(\bar v_1+\lambda(E)D^2)\bmod Q$.
\State Apply the diagonal unitary $U_{\mathrm{corr}}^{(v_{1,Q})}$ on $X_1$ as defined in AC4.
\State Apply $\mathrm{QFT}_{\ZZ_M}^{\otimes n}$ to $X$ and measure $\bm u\in\ZZ_M^n$.
\State \textbf{return} $\bm u_Q\gets \bm u\bmod Q$.
\end{algorithmic}
\end{algorithm}

\section{The new Step \texorpdfstring{$9^{\dagger}$}{9 dagger}}
\label{sec:step9dagger}

\begin{algorithm}[H]
\caption{Step $9^{\dagger}$}
\label{algo:1}
\begin{algorithmic}[1]
\State \textbf{Input:} A chosen-secret LWE instance $(\mathbf U,\mathbf t)$ as in~\cite{chen2024quantum} over modulus $q$, equivalently the lattice $L_q^\perp(\mathbf A)$ with $\mathbf A=[2p_1\mathbf t\mid \mathbf U^{\mathsf T}\mid \mathbf I_m]$, together with public parameters $(q,\beta,p_1,\dots,p_\kappa)$.
\State \textbf{Oracle:} State-preparation access to $|\tilde{\varphi}_7\rangle$ as in Definition~\ref{def:hidden-slope}.
\State \textbf{Parameters:} A re-tuned Karst-wave configuration $(c,D,r,s,t,u,P,M,Q)$ as in Definition~\ref{def:retuned-karst-wave-config}.
\State \textbf{Output:} Either the LWE secret $s$ and error $e$, or \textsc{fail}.
\State Let $\ell\gets$ the number of rows of $\mathbf U$ and $m\gets$ the number of columns of $\mathbf U$; set $n\gets 1+\ell+m$.
\State Fix absolute constants $C_{\mathrm{rank}}>1$, $C_{\mathrm{try}}\ge 1$, and $C_{\mathrm{samp}}\ge 1$.
\State Set the target sample count $N\gets \lceil C_{\mathrm{rank}}(n-1)\rceil$ and the batch budget $T_{\max}\gets \lceil C_{\mathrm{try}}n\rceil$.
\State Set a per-batch oracle-call cap $B_{\max}\gets \lceil C_{\mathrm{samp}}\cdot N\rceil$.
\State Precompute $\mathrm{inv}:=(-2D^2)^{-1}\in\ZZ_Q$.
\For{$\mathrm{trial}=1$ \textbf{to} $T_{\max}$}
\State \quad Initialize $\mathcal{S}\gets\emptyset$ and $\mathrm{calls}\gets 0$.
\While{$|\mathcal{S}|<N$ \textbf{and} $\mathrm{calls}<B_{\max}$}
\State \quad\quad $\mathrm{calls}\gets \mathrm{calls}+1$.
\State \quad\quad $\bm u_Q \gets \textsc{Step9DaggerSample}(D,M,Q,\mathrm{inv})$.
\If{$\bm u_Q\neq \bm 0$}
\State \quad\quad\quad Append $\bm u_Q$ as a row of $\mathcal{S}$.
\EndIf
\EndWhile
\If{$|\mathcal{S}|<N$}
\State \quad\quad \textbf{continue}
\EndIf
\State \quad $\bbb_Q \gets \textsc{RecoverBModQ}(\mathcal{S},Q)$.
\If{$\bbb_Q=\textsc{fail}$}
\State \quad\quad \textbf{continue}
\EndIf
\State \quad $(s,e) \gets \textsc{ExtractAndVerifyLWE}(\bbb_Q,\mathbf U,\mathbf t,q,\beta,p_1,\dots,p_\kappa)$.
\If{$(s,e)\neq\textsc{fail}$}
\State \quad\quad \textbf{return} $(s,e)$.
\EndIf
\EndFor
\State \textbf{return} \textsc{fail}.
\end{algorithmic}
\end{algorithm}

\begin{algorithm}[H]
\caption{\textsc{ExtractAndVerifyLWE}: parse and certify $(s,e)$}
\label{algo:extract-lwe}
\begin{algorithmic}[1]
\State \textbf{Input:} $\bbb_Q\in\ZZ_Q^n$, a chosen-secret LWE instance $(\mathbf U,\mathbf t)$ with $\mathbf U\in\ZZ_q^{\ell\times m}$ and $\mathbf t\in\ZZ_q^{m}$, and public parameters $(q,\beta,p_1,\dots,p_\kappa)$.
\State \textbf{Output:} The LWE secret $s$ and error $e$, or \textsc{fail}.
\State Let $\ell\gets$ the number of rows of $\mathbf U$ and $m\gets$ the number of columns of $\mathbf U$; set $n\gets 1+\ell+m$.
\If{$\kappa-1>\ell$}
\State \textbf{return} \textsc{fail}.
\EndIf
\State Lift $\bbb_Q$ coordinate-wise to $\bbb\in(-Q/2,Q/2]^n\cap\ZZ^n$.
\If{$b_1\neq -1$ as an integer}
\State \textbf{return} \textsc{fail}.
\EndIf
\For{$i=2$ \textbf{to} $n$}
\If{$2p_1\nmid b_i$ in $\ZZ$}
\State \textbf{return} \textsc{fail}.
\EndIf
\EndFor
\For{$i=2$ \textbf{to} $\kappa$}
\If{$b_i \neq 2p_1p_i$ in $\ZZ$}
\State \textbf{return} \textsc{fail}.
\EndIf
\EndFor
\State Parse $\bbb=[-1,\,2p_1 s^\top,\,2p_1 e^\top]^\top$ with $s\in\ZZ^\ell$ and $e\in\ZZ^m$, and recover $s,e$ by integer division by $2p_1$.
\State Check shortness, e.g.\ $\|s\|_\infty,\|e\|_\infty\le \beta\log n$, and any promised norm bounds for $\|\bbb\|$ from the lattice analysis.
\If{$\mathbf U^{\mathsf T}s+e\not\equiv \mathbf t\pmod q$}
\State \textbf{return} \textsc{fail}.
\EndIf
\State \textbf{return} $(s,e)$.
\end{algorithmic}
\end{algorithm}

\begin{algorithm}[H]
\caption{\textsc{RecoverBModQ}: solve for $\bbb\bmod Q$ from dual samples}
\label{algo:recover-bmodQ}
\begin{algorithmic}[1]
\State \textbf{Input:} A matrix $\mathcal{S}\in(\ZZ_Q)^{N\times n}$ whose rows are nonzero samples $\bm u_Q$.
\State \textbf{Output:} $\bbb_Q=\bbb\bmod Q$, or \textsc{fail}.
\State Let $W\in(\ZZ_Q)^{N\times(n-1)}$ be the submatrix of $\mathcal{S}$ consisting of columns $2..n$, and let $\bm y\in(\ZZ_Q)^N$ be column $1$ of $\mathcal{S}$.
\State Solve $W\bm z\equiv \bm y\pmod Q$ and test uniqueness, e.g.\ via Smith normal form~\citep{kannan1979polynomial}.
\If{no solution exists \textbf{or} the solution is not unique}
\State \textbf{return} \textsc{fail}.
\EndIf
\State Let $\bm z$ be the unique solution and set $\bbb_Q\gets[-1,\bm z^\top]^\top\in\ZZ_Q^n$.
\State \textbf{return} $\bbb_Q$.
\end{algorithmic}
\end{algorithm}

\begin{remark}[Classical guessing loop for the norm]\label{rem:guess-loop}
As in~\cite{chen2024quantum}, the small modulus is chosen as $Q=(c{+}1)\|\bbb\|^{2}$, where $\bbb$ is the unknown unique shortest vector of $L_q^\perp(\mathbf A)$.
We run a classical outer loop over all polynomially many candidates $B_{\mathrm{guess}}$ for $\|\bbb\|^{2}$.
For each guess we set $Q\gets(c{+}1)B_{\mathrm{guess}}$, choose an odd $D$ with $\gcd(D,Q)=1$, derive $M=2D^{2}Q$ and $M_2=D^{2}Q$, and complete the remaining Karst-wave parameters as in Definition~\ref{def:retuned-karst-wave-config}.
Guesses for which $Q$ is not odd, or for which the remaining re-tuned parameters cannot be chosen to satisfy AC2--AC5, are skipped.
We then execute Algorithm~\ref{algo:1} for this guess.

Only the correct guess is guaranteed by the analysis to yield the dual samples used in Theorem~\ref{thm:correct}.
An incorrect guess may cause sampling or rank recovery to fail; if it nevertheless returns a pair $(s,e)$, the explicit LWE verification in Algorithm~\ref{algo:extract-lwe} ensures that the returned pair is correct.
To keep rejection for incorrect guesses polynomial-time, Algorithm~\ref{algo:1} uses a fixed trial budget $T_{\max}=\poly(n)$ and a per-trial sampling cap $B_{\max}=\poly(n)$, returning \textsc{fail} if no verified solution is found within budget.
\end{remark}

\begin{remark}[Instantiation for LWE]
For the $q$-ary lattice $L_q^\perp(\mathbf A)$ in~\citet{chen2024quantum}, the promised unique shortest vector has the form
\[
\bbb=[-1,\,2p_1 s^\top,\,2p_1 e^\top]^\top\in\mathbb Z^n,
\]
where $s$ is the LWE secret and $e$ is the error vector.
Because $\|\bbb\|_\infty \le 2p_1\beta\log n < Q/2$ in the parameter regime of interest, the reduction map $\bbb\mapsto \bbb\bmod Q$ is injective on the centered cube that contains the true shortest vector.
Therefore, once Algorithm~\ref{algo:1} recovers $\bbb\bmod Q$, the canonical lift to $(-Q/2,Q/2]^n$ recovers $\bbb$ as an integer vector.
Dividing the last $n-1$ coordinates by $2p_1$, after checking divisibility, yields $s$ and $e$, and the check
$\mathbf U^{\mathsf T}s+e\equiv \mathbf t\pmod q$ certifies correctness.
\end{remark}

\paragraph{Fourier sampling.}
We apply $\mathrm{QFT}_{\ZZ_{M}}^{\otimes n}$ to the coordinate register block and then measure $\bm u\in\ZZ_{M}^n$.
The outcome distribution is analyzed next.

\subsection{Correctness under Additional Conditions}

\begin{lemma}[Dual sampling via direct QFT]\label{lem:orth-exact}
Assume AC1--AC5 and let $\ket{\varphi_7}$ be the chirp-free state from Lemma~\ref{lem:chirp-cancellation}.
Fix a good measurement transcript $E$ in the sense of Additional Condition~AC5 and write the conditional state as
\[
\ket{\varphi_7}
= \sum_{j\in\ZZ}\sum_{\bm k\in \calK} \alpha_E(j,\bm k)\,i^{\|\bm k\|^2}\,
\ket{j\Delta + \bm v^\ast + M_2 \bm k \bmod M},
\]
where $\Delta = 2D^2\bbb$ and $M_2 = M/2$.
After applying $\mathrm{QFT}_{\ZZ_{M}}^{\otimes n}$ to the coordinate registers and measuring $\bm u \in \ZZ_{M}^n$, we have
\[
\Pr\bigl[\ip{\bbb}{\bm u}\equiv 0 \pmod Q\bigr] \ge 1 - \negl(n).
\]
Moreover, by Lemma~\ref{lem:uniform-modQ}, conditioned on the event $\ip{\bbb}{\bm u}\equiv 0\pmod Q$ the reduced outcome
$\bm u\bmod Q$ is exactly uniform on the dual hyperplane $H$, hence
\[
\Pr[\bm u\bmod Q=\bm 0\mid \ip{\bbb}{\bm u}\equiv0\pmod Q]= Q^{-(n-1)}.
\]
\end{lemma}

\begin{proof}
Fix $E$.
Apply $\mathrm{QFT}_{\ZZ_M}^{\otimes n}$ to the chirp-free state~\eqref{eq:phi-intro}.
For each $\bm u\in\ZZ_M^n$, the unnormalized amplitude equals
\[
\omega_M^{\ip{\bm u}{\bm v^\ast}}\cdot
\sum_{j\in\ZZ}\sum_{\bm k\in\calK}\alpha_E(j,\bm k)\,i^{\|\bm k\|^2}\,
\omega_M^{\,j\ip{\bm u}{\Delta}+M_2\ip{\bm u}{\bm k}},
\]
so the outcome probabilities are proportional to the squared magnitude of the inner sum.

By AC5, conditioned on $E$ the coefficients factor as $\alpha_E(j,\bm k)=\alpha_E(j)\beta_E(\bm k)$.
Using $M_2=M/2$ we have $\omega_M^{M_2\ip{\bm u}{\bm k}}=(-1)^{\ip{\bm u}{\bm k}}$, and since $k_1=0$ this depends only on the parity pattern of $\bm u_{[2..n]}$.
Moreover $M=2D^2Q$ and $\Delta=2D^2\bbb$, hence
\[
\omega_M^{\,j\ip{\bm u}{\Delta}}
=\exp\!\left(2\pi i\,j\,\frac{t(\bm u)}{Q}\right),
\]
where $t(\bm u):=\ip{\bbb}{\bm u}\bmod Q\in\ZZ_Q$ and we identify $t(\bm u)$ with its representative in $\{0,\dots,Q-1\}$ when used inside the exponential.
Equivalently, if we write $\bm u=\bm u_Q+Q\bm t'$ with $\bm u_Q=\bm u\bmod Q$, then
\[
\ip{\bm u}{\Delta}-\ip{\bm u_Q}{\Delta}=\ip{Q\bm t'}{2D^2\bbb}\in M\ZZ,
\]
so $\omega_M^{j\ip{\bm u}{\Delta}}=\omega_M^{j\ip{\bm u_Q}{\Delta}}$ for all integers $j$.
Therefore $S_E(\bm u)$ factorizes as
\[
S_E(\bm u)=
\Bigl(\sum_{j\in\ZZ}\alpha_E(j)\,e^{2\pi i j\,t(\bm u)/Q}\Bigr)\cdot
\Bigl(\sum_{\bm k\in\calK}\beta_E(\bm k)\,i^{\|\bm k\|^2}\,(-1)^{\ip{\bm u}{\bm k}}\Bigr)
= \wh\alpha_E(t(\bm u)/Q)\cdot G_E(\bm u),
\]
where $G_E(\bm u)$ depends only on the parity pattern of $\bm u_{[2..n]}$.
Thus
\[
\Pr[\bm u]\ \propto\ |S_E(\bm u)|^2
=|\wh\alpha_E(t(\bm u)/Q)|^2\cdot |G_E(\bm u)|^2.
\]

For each residue $a\in\ZZ_Q$, let
\[
T_a:=\{\bm u\in\ZZ_M^n:\ip{\bbb}{\bm u}\equiv a\pmod Q\}.
\]
Because $b_1=-1$ is a unit in $\ZZ_Q$, for every fixed $\bm u_{[2..n]}\in\ZZ_M^{n-1}$ and every $a\in\ZZ_Q$ there are exactly $M/Q=2D^2$ values of $u_1\in\ZZ_M$ such that $\bm u\in T_a$.
Since $G_E(\bm u)$ does not depend on $u_1$, the total grid mass is independent of $a$.
Writing $d:=n-1$ and
\[
B_E(\rho):=
\sum_{\bm k\in\calK}
\beta_E(\bm k)i^{\|\bm k\|^2}
(-1)^{\ip{\rho}{\bm k_{[2..n]}}},
\qquad \rho\in\{0,1\}^{d},
\]
we have $G_E(\bm u)=B_E(\bm u_{[2..n]}\bmod2)$ and hence
\[
\sum_{\bm u\in T_a}|G_E(\bm u)|^2
=
\frac{M}{Q}\left(\frac M2\right)^d
\sum_{\rho\in\{0,1\}^d}|B_E(\rho)|^2
=: C_E.
\]
The constant $C_E$ is positive for a nonzero factorized state.
Indeed, by orthogonality of the characters
$\rho\mapsto(-1)^{\ip{\rho}{\bm k_{[2..n]}}}$ on $\{0,1\}^{d}$,
\[
\sum_{\rho\in\{0,1\}^d}|B_E(\rho)|^2
=2^d\sum_{\bm k\in\calK}|\beta_E(\bm k)|^2>0.
\]
Therefore
\[
\Pr[T_a]
=
\frac{|\wh\alpha_E(a/Q)|^2}{\sum_{s\in\ZZ_Q}|\wh\alpha_E(s/Q)|^2}.
\]

By the leakage bound~\eqref{eq:AC5-leakage} and this proportionality, the total off-resonant probability mass satisfies
\[
\sum_{a\neq 0}\Pr[T_a]
=
\frac{\sum_{a\neq 0}|\wh\alpha_E(a/Q)|^2}{\sum_{s\in\ZZ_Q}|\wh\alpha_E(s/Q)|^2}
\le \negl(n),
\]
so
\[
\Pr\bigl[\ip{\bbb}{\bm u}\equiv 0\pmod Q\bigr]\ge 1-\negl(n).
\]
Finally, conditioned on this resonance event, Lemma~\ref{lem:uniform-modQ} gives exact uniformity of $\bm u\bmod Q$ on $H$, and in particular
\[
\Pr[\bm u\bmod Q=\bm 0\mid \ip{\bbb}{\bm u}\equiv0\pmod Q]=Q^{-(n-1)}.
\]
\end{proof}

\begin{theorem}[Step $9^{\dagger}$ is correct]\label{thm:correct}
Assume AC1--AC5 and run Algorithm~\ref{algo:1} in the correct iteration of the outer norm-guessing loop.
Each accepted sample $\bm u_Q\in\ZZ_Q^n$ appended to $\mathcal S$ by Algorithm~\ref{algo:1} is non-zero by construction and satisfies
\[
\ip{\bbb}{\bm u_Q} \equiv 0 \pmod{Q}
\]
except with total probability $\negl(n)$ over the transcript distribution and the internal randomness of the measurements.
With overwhelming probability, the accepted resonant samples give a unique solution to the recovery system modulo $Q$.
This solution is $\bbb\bmod Q$, and its centered lift is the integer vector $\bbb$.
The LWE secret and error are then extracted and verified in polynomial time with overwhelming probability.
\end{theorem}

\begin{proof}
By Additional Condition~AC5, a transcript drawn from the front end is good except with probability $\negl(n)$.
Algorithm~\ref{algo:1} makes only $\poly(n)$ oracle calls, so by a union bound we may condition on the event that all invocations are good, losing at most a negligible term in the overall success probability.

In each oracle call within Algorithm~\ref{algo:1}, the sampling primitive \textsc{Step9DaggerSample} measures $\tau=X_1\bmod D^2$, obtains the run-local residue $v_{1,Q}=v_1^\ast(E)\bmod Q$ from run-local side information, and applies the diagonal gate $U_{\mathrm{corr}}^{(v_{1,Q})}$.
Lemma~\ref{lem:chirp-cancellation} shows that this removes the quadratic chirp and produces the chirp-free state $|\varphi_7\rangle$ assumed in AC5.

Applying $\mathrm{QFT}_{\ZZ_M}^{\otimes n}$ and measuring yields $\bm u\in\ZZ_M^n$.
Lemma~\ref{lem:orth-exact} implies that $\langle \bbb,\bm u\rangle\equiv 0\pmod Q$ except with probability $\negl(n)$, and Lemma~\ref{lem:uniform-modQ} implies that, conditioned on this resonance event, $\bm u\bmod Q$ is exactly uniform on
\[
H=\{\bm v\in\ZZ_Q^n:\langle \bbb,\bm v\rangle\equiv 0\pmod Q\}.
\]
In particular,
\[
\Pr[\bm u\bmod Q=\bm 0]\le Q^{-(n-1)}+\negl(n).
\]
Let $d:=n-1$.
Since $Q\ge5$ in the re-tuned regime, $Q^{-d}=\negl(n)$.
Because $B_{\max}\ge N$, it is enough to look at the first $N$ oracle calls in a batch.
A union bound shows that, with probability $1-\negl(n)$, all of these first $N$ calls are nonzero and resonant; hence the batch obtains $N$ accepted samples before the cap is reached, and every accepted $\bm u_Q$ satisfies
$\langle \bbb,\bm u_Q\rangle\equiv 0\pmod Q$.
We condition on this high-probability event for the rank argument below; removing the conditioning changes the final success probability by only $\negl(n)$.

Fix such a successful batch.
Because $b_1=-1$ is a unit in $\ZZ_Q$, the projection $H\to\ZZ_Q^{n-1}$ given by $\bm u\mapsto \bm u_{[2..n]}$ is a bijection.
Moreover, on $H$ the only vector with $\bm u_{[2..n]}=\bm 0$ is $\bm u=\bm 0$, because $-u_1\equiv 0$ then forces $u_1\equiv 0$.
Thus rejecting $\bm u_Q=\bm 0$ is equivalent, on resonance, to rejecting $\bm u_{Q,[2..n]}=\bm 0$.
Under the conditioning event, the accepted resonant vectors $\bm u_{Q,[2..n]}$ are independent and exactly uniform over $\ZZ_Q^{d}\setminus\{\bm 0\}$.
Let $W\in\ZZ_Q^{N\times d}$ be the matrix with these rows and let $\bm y$ be the vector of first coordinates.
Then the true unknown $\bbb_{[2..n]}\bmod Q$ satisfies $W\bbb_{[2..n]}\equiv \bm y\pmod Q$.

It remains to justify uniqueness over the composite ring $\ZZ_Q$.
For a prime $p\mid Q$, consider one accepted tail row reduced modulo $p$.
Under the exact distribution uniform on $\ZZ_Q^d\setminus\{\bm 0\}$, its reduction to $\mathbb F_p^d$ has probabilities
\[
\Pr[\bm a]
=
\begin{cases}
\dfrac{(Q/p)^d-1}{Q^d-1}, & \bm a=\bm 0\in\mathbb F_p^d,\\[2ex]
\dfrac{(Q/p)^d}{Q^d-1}, & \bm a\neq\bm 0\in\mathbb F_p^d.
\end{cases}
\]
Thus its total-variation distance from the uniform distribution on $\mathbb F_p^d$ is at most
\[
\frac{1-p^{-d}}{Q^d-1}\le \negl(n).
\]
By tensorization, the joint law of the $N$ reduced rows is within $N\negl(n)=\negl(n)$ of the law of an i.i.d.\ uniform $N\times d$ matrix over $\mathbb F_p$.
For an i.i.d.\ uniform matrix $A\in\mathbb F_p^{N\times d}$,
\[
\Pr[\operatorname{rank}_{\mathbb F_p}(A)<d]
\le
\Pr[\exists \bm z\neq 0: A\bm z=0]
\le
(p^d-1)p^{-N}
\le p^{d-N}
=p^{-\Omega(n)},
\]
because $N=\lceil C_{\mathrm{rank}}d\rceil$ with $C_{\mathrm{rank}}>1$.
Since $Q=\poly(n)$, the number of distinct prime divisors of $Q$ is at most $\log_2 Q=O(\log n)$, so the preceding total-variation bound and a union bound over all $p\mid Q$ give full column rank modulo every prime divisor of $Q$ with probability $1-\negl(n)$.

If $\operatorname{rank}_{\mathbb F_p}(W\bmod p)=d$ and $W\bm z\equiv \bm 0\pmod{p^e}$, then reducing modulo $p$ gives $\bm z\equiv \bm 0\pmod p$.
Writing $\bm z=p\bm z_1$ modulo $p^e$, the congruence $W\bm z\equiv0\pmod{p^e}$ implies $W\bm z_1\equiv0\pmod{p^{e-1}}$.
Repeating the same argument shows $\bm z\equiv \bm0\pmod{p^e}$.
Thus $\ker(W\bmod p^e)=\{\bm 0\}$.
By the Chinese remainder theorem, $\ker(W\bmod Q)=\{\bm 0\}$.
Therefore the system has a unique solution $\bm z\in\ZZ_Q^{n-1}$, which Algorithm~\ref{algo:recover-bmodQ} computes and uses to set
\[
\bbb_Q=[-1,\bm z^\top]^\top=\bbb\bmod Q.
\]

Finally, the bound $\|\bbb\|_\infty<Q/2$ from (R1) makes the centered lift from $\ZZ_Q^n$ to $(-Q/2,Q/2]^n$ injective on the relevant region, so lifting $\bbb_Q$ recovers $\bbb$ as an integer vector.
The subsequent divisibility, planted-prefix, shortness, and LWE-consistency checks in Algorithm~\ref{algo:extract-lwe} succeed for the correct guess.
For any other guess, the algorithm outputs only if these explicit checks certify a valid LWE solution.
\end{proof}

\begin{corollary}[Conditional quantum algorithm for LWE]\label{cor:lwe}
Assume the LWE-to-lattice reduction and the unique-shortest-vector promise for $L_q^\perp(\mathbf A)$ from~\citet{chen2024quantum}.
If the corresponding front end can be compiled in a re-tuned configuration satisfying AC1--AC4 and the Gaussian-envelope hypotheses of Proposition~\ref{prop:AC5-holds}, with perturbations controlled for example by Lemma~\ref{lem:AC5-perturbation}, then Proposition~\ref{prop:AC5-holds} supplies AC5, and Algorithm~\ref{algo:1} solves the chosen-secret LWE instance encoded by $L_q^\perp(\mathbf A)$ in quantum polynomial time with overwhelming probability.
Via the classical reductions of~\citet{chen2024quantum}, this yields a quantum algorithm for standard LWE in the same parameter regime, conditional on AC1--AC4 and the Gaussian-envelope hypotheses that imply AC5.
\end{corollary}

\section{Related Work}
\label{sec:related-work}

Lattice-based cryptography is rooted in worst-case hardness frameworks such as Ajtai's construction of hard lattice instances~\citep{ajtai1996generating} and subsequent worst-case-to-average-case reductions based on Gaussian measure~\citep{micciancio2007worst}.
Regev introduced LWE and gave quantum worst-case reductions from lattice problems such as GapSVP and SIVP to LWE~\citep{regev2009lattices}.
Subsequent work clarified parameter tradeoffs and established classical hardness in a range of regimes~\citep{peikert2009public,brakerski2013classical}.
These reductions underpin a broad ecosystem of cryptographic constructions, including trapdoors for hard lattices~\citep{gentry2008trapdoors} and fully homomorphic encryption~\citep{gentry2009fully}.

For context, the fastest known worst-case SVP algorithms are classical and typically rely on sieving and/or discrete Gaussians, starting from the sieve of Ajtai--Kumar--Sivakumar~\citep{ajtai2001sieve} and later practical refinements~\citep{nguyen2008sieve}, and including the $2^{n+o(n)}$-time discrete-Gaussian algorithm of~\citet{aggarwal2015solving}.
Discrete Gaussian sampling is also a central technical tool in lattice algorithms more broadly; efficient lattice Gaussian samplers were developed, for example, by Peikert~\citep{peikert2010efficient}.

On the quantum side, lattice algorithms are often organized around preparing structured coset- or line-supported superpositions and extracting information via Fourier sampling.
A canonical early connection is Regev's reduction from unique-SVP to the dihedral hidden subgroup problem via coset sampling~\citep{regev2004quantum}, together with subexponential-time algorithms for the dihedral hidden subgroup problem such as Kuperberg's sieve~\citep{kuperberg2005subexponential}.
Related dihedral-coset models have since appeared in lattice settings, e.g., the extrapolated dihedral coset problem, which is closely connected to LWE~\citep{brakerski2018learning}.

On the average-case state side, Chen, Liu, and Zhandry introduced a filtering framework for decoding hidden linear structure from LWE-like quantum states and used it to obtain polynomial-time quantum algorithms for several average-case variants, including EDCP in certain regimes~\citep{chen2022quantum}.
Subsequent work has studied both algorithms and hardness barriers for quantum LWE-state families with structured, e.g.\ Gaussian, amplitudes and linear or quadratic phase terms~\citep{chen2025lwe}.

Chen's Karst-wave construction~\citep{chen2024quantum} is an instance of this Fourier-sampling template with amplitude engineering via complex-Gaussian windowing, yielding a post-processing state supported on a shifted grid line whose amplitudes include a quadratic chirp in the loop index.
Chen's published post-processing (Steps~8--9) exploits additional arithmetic structure to handle the unknown shift and chirp.
The present work isolates an explicit access-model promise that suffices for exact post-processing in the LWE regime:
run-local access to $v_{1,Q}:=v_1^\ast(E)\bmod Q$ (AC4), enabling single-coordinate chirp cancellation followed by a direct $\mathrm{QFT}_{\ZZ_M}^{\otimes n}$ without learning the full offset $\bm v^\ast(E)$.
For Gaussian loop envelopes, the needed resonance condition AC5 follows from one-dimensional Fourier decay on the $Q$-point grid.

\section{Conclusion}

We present a new Step~$9^{\dagger}$ that turns unknown affine offsets into harmless Fourier phases.
Under AC1--AC4 and the Gaussian spectral hypotheses that imply AC5, it yields modular linear relations from direct Fourier sampling on the coordinate registers.
The key post-processing ingredients are run-local chirp cancellation from the first coordinate and exact reduction of resonant Fourier samples modulo $Q$.
The reduced resonant samples are exactly uniform on the dual hyperplane, which makes the final linear-algebraic recovery over the composite ring $\ZZ_Q$ straightforward.

We describe the corresponding conditional instantiation with the parameters and the lattice $L_q^\perp(\mathbf A)$ from~\citet{chen2024quantum}.
In the access model of Section~\ref{subsec:access-model}, such a construction gives a direct path from a superposition with support $\{j\Delta+\bm v^\ast\}$ to samples from the dual lattice.
Under AC1--AC4 and the Gaussian-envelope hypotheses proved above to imply AC5, the resulting algorithm solves the chosen-secret LWE problem of~\cite{chen2024quantum} in the same qualitative parameter regime.

\section*{Acknowledgments}

We are grateful to all who provided constructive discussions and helpful feedback.
The author used AI-enabled tools solely for English grammar and clarity suggestions in non-technical prose.
All technical ideas, proofs, and results are the author's own work.

\vspace{5ex}
\bibliography{reference}
\bibliographystyle{plainnat}





\end{document}